\documentclass{nature}

\usepackage{braket}
\usepackage{graphicx}
\usepackage{amsmath}

\newcommand{\means}[1]{\langle#1\rangle}
\usepackage{bm}

\title{Direct Evidence for Fermi Statistics from Proximity to the Kitaev Spin Liquid in $\alpha$-RuCl$_{3}$}

\author{Yiping Wang,$^{1}$ Gavin B. Osterhoudt,$^{1}$ Yao Tian,$^{2}$ Paige Lampen-Kelley,$^{3}$ Arnab Banerjee,$^{4}$ Thomas Goldstein,$^{5}$ Jun Yan,$^{5}$ Johannes Knolle,$^{6}$ Huiwen Ji,$^{7}$, Robert J. Cava,$^{7}$ Joji Nasu,$^{8}$ Yukitoshi Motome,$^{9}$ Stephen E. Nagler,$^{4}$ David Mandrus,$^{3}$ Kenneth S. Burch$^{1}$}

\begin{document}
\maketitle

\begin{affiliations}
 \item Department of Physics, Boston College, 140 Commonwealth Avenue, Chestnut Hill, MA 02467, USA
 \item SICK Product Center Asia Pte. Ltd., 8 Admiralty Street, Singapore 757438
 \item Department of Materials Science and Engineering, University of Tennessee, Knoxville, TN 37821, USA
 \item Neutron Scattering Division, Oak Ridge National Laboratory, Oak Ridge, TN 37831, USA
 \item Department of Physics, University of Massachusetts, Amherst, Massachusetts 01003, USA
 \item Blackett Laboratory, Imperial College London, London SW7 2AZ, United Kingdom
 \item Department of Chemistry, Princeton University, Princeton, NJ 08540, USA 
 \item Department of Physics, Yokohama National University, Hodogaya, Yokohama 240-8501, Japan
 \item Department of Applied Physics, University of Tokyo, Bunkyo, Tokyo 113-8656, Japan
\end{affiliations}

\begin{abstract}
A key characteristic of quantum spin liquids(QSL) is the presence of fractional excitations related to their entanglement properties, yet  experimental verification of their statistics is missing.\cite{Wen2016,Matern2017,Savary2017,Kitaev2006} For example, in the potential Kitaev spin liquid, $\alpha$-RuCl$_3$, experiments uncovered signs of fractional particles,\cite{Banerjee2016,Kasahara2018,Koitzsch2016,WangLoidl2017,Winter2017,Zheng2017,Wellm2018,Do2017} though not their Fermi statistics. Here we employ Raman scattering to reveal the true nature of the magnetic excitations, using improved experimental methods and analysis to remove the influence of laser heating and thermal fluctuations. Via the energy loss and gain spectra, we extract the energy and temperature dependence of the Raman susceptibility to prove $\alpha$-RuCl$_{3}$'s magnetic response is given by pair creation of fermionic excitations. Furthermore, by comparing with quantum Monte Carlo (QMC) results for the exact Kitaev limit, we are able to discern the energy and temperature range where additional interaction terms are important. Our results open new directions in QSL research by providing a new way to investigate fractional excitations and the importance of terms causing spinon confinement. 
\end{abstract}

Correlated materials can serve as new vacua out of which novel particles appear as collective modes. For example, in magnetic insulators the breaking of time reversal symmetry below the ordering temperature leads to the emergence of magnons. Qualitatively distinct are fractionalized excitations in frustrated magnets that cannot be understood as simple local combinations of quantum numbers of the materials' constituents. These have long been sought after in quantum spin liquids (QSL) where long-range entanglement and non-trivial topology produce elementary excitations that are non-local in terms of spin flips\cite{Wen2016,Matern2017,Savary2017,Kitaev2006}. QSL excitations come in a rich variety, e.g. fermions, bosons, or even anyons, with elevated energy and temperature scales that make these particles promising building blocks for fault tolerant topological quantum computers\cite{Kitaev2006}. However, establishing that the excitations are indeed fractional in QSL, has been a long standing challenge. In part, the difficulty arises from the need to separate the response from other contributions (e.g. phonons) to clearly reveal their fermionic or anyonic statistics. This has been further compounded by the lack of exactly solvable models that can be directly compared with experiments and methods that uncover the effects of additional terms relieving frustration. 

Attention has focused on relativistic Mott insulators that could realize the exactly solvable Kitaev model. In materials such as AIrO$_{3}$\cite{singh_antiferromagnetic_2010,2010PhRvL.105b7204C,Kitagawa2018,abramchuk_cu2iro3:_2017} and $\alpha$-RuCl$_{3}$\cite{PLumb2014}, the large spin-orbit coupling and Coulomb repulsion result in j$_{eff}=1/2$ moments on the honeycomb lattice. Similar to Kitaev's original proposal,\cite{Kitaev2006} these result in an Ising type interaction whose component depends on the particular bond connecting the two effective spins. The resulting excitations, created by spin flips, are Z2 gauge fluxes and dispersive Majorana fermions. However, the presence of additional Heisenberg and other bond-dependent terms typically leads to magnetic order in these compounds.\cite{Winter2017,Catuneanu2018} In $\alpha$-RuCl$_{3}$, the application of a magnetic field destroys the antiferromagnetic order and there is mounting evidence for proximity to a quantum spin liquid state.\cite{Zheng2017,Kasahara2018} Nonetheless, whether the excitations have fractional statistics and over what energy and temperature range has not been established. 

Dynamical scattering experiments can confirm the presence of fractional particles in QSL by probing their creation and annihilation, dictated by their kinematics and statistics.\cite{Sandilands2015,Banerjee2016}  Similar to asymptotic freedom in quantum chromodynamics, above the magnetic ordering temperature fractionalized excitations may dominate the short time (higher frequency) behavior of materials with a nearby QSL phase.\cite{Banerjee2016} Raman scattering is extremely useful in this regard, as it can study magnetic excitations' symmetry and statistics\cite{devereaux:175} with high energy and temperature resolution, even in single 2D atomic layers.\cite{Kim2019,BurchNatMag18} Indeed, the Raman susceptibility could confirm the fractional nature of the excitations via the temperature and energy dependence of their mutual statistics (i.e. being governed by Fermi functions). Raman scattering was the first to reveal the continuum from magnetic excitations in $\alpha$-RuCl$_{3}$,\cite{Sandilands2015} and the temperature dependence was invoked to suggest the excitations are fermionic\cite{Nasu2016,Glamazda2017}. While the presence of a continuum in $\alpha$-RuCl$_{3}$ is well established,\cite{WangLoidl2017,Little2017,Banerjee2016,Zheng2017,Do2017} it has been argued that the continuum may be a result of strongly interacting magnons\cite{Winter2017} rather than fractional excitations. Furthermore the role of the non-Kitaev terms remains controversial. In particular, previous Raman efforts have been plagued by laser heating\cite{Sandilands2015} from the large incident power required to observe the low intensity continuum. While others claimed a fermionic response at room temperature due to a lack of clear separation of the bosonic contribution\cite{Glamazda2017}. Moreover, these focused strictly on the integrated Raman intensity that was contaminated by a Bose factor.

We overcome these previous limitations with new anti-Stokes spectra and a unique framework to prove the fractional nature of the excitations in $\alpha$-RuCl$_{3}$ via their statistical properties. With enhanced signals we could minimze the laser power and ensure the absence of unwanted heating via both energy gain as well as loss measurements. These are then combined to reveal the Raman susceptibility, where the continuum is unambiguously separated from bosonic thermal fluctuations. As such, the Raman susceptibility provides direct evidence that the energy and temperature dependence of the magnetic excitations in $\alpha$-RuCl$_3$ are governed by Fermi statistics. Surprisingly, despite the presence of other terms and high temperatures relative to the flux gap, we find good agreement with the simple prediction of fractional particles governed by Fermi statistics. Using this result we also extract the joint density of states and find it is nearly independent of temperature as well as energy, consistent with numerical calculations.\cite{PhysRevB.92.115122} Direct comparison with QMC calculations for the Kitaev model reveals the very low energy and temperature response originates from additional terms in  $\alpha$-RuCl$_{3}$. Thus our new approach solidifies the wide energy and temperature range over which the magnetic excitations in $\alpha$-RuCl$_{3}$ are fermionic. Furthermore our procedure can be used to identify the fractional nature and nontrivial statistics of QSLs.

\begin{figure}
\centering
\includegraphics[width=1\textwidth]{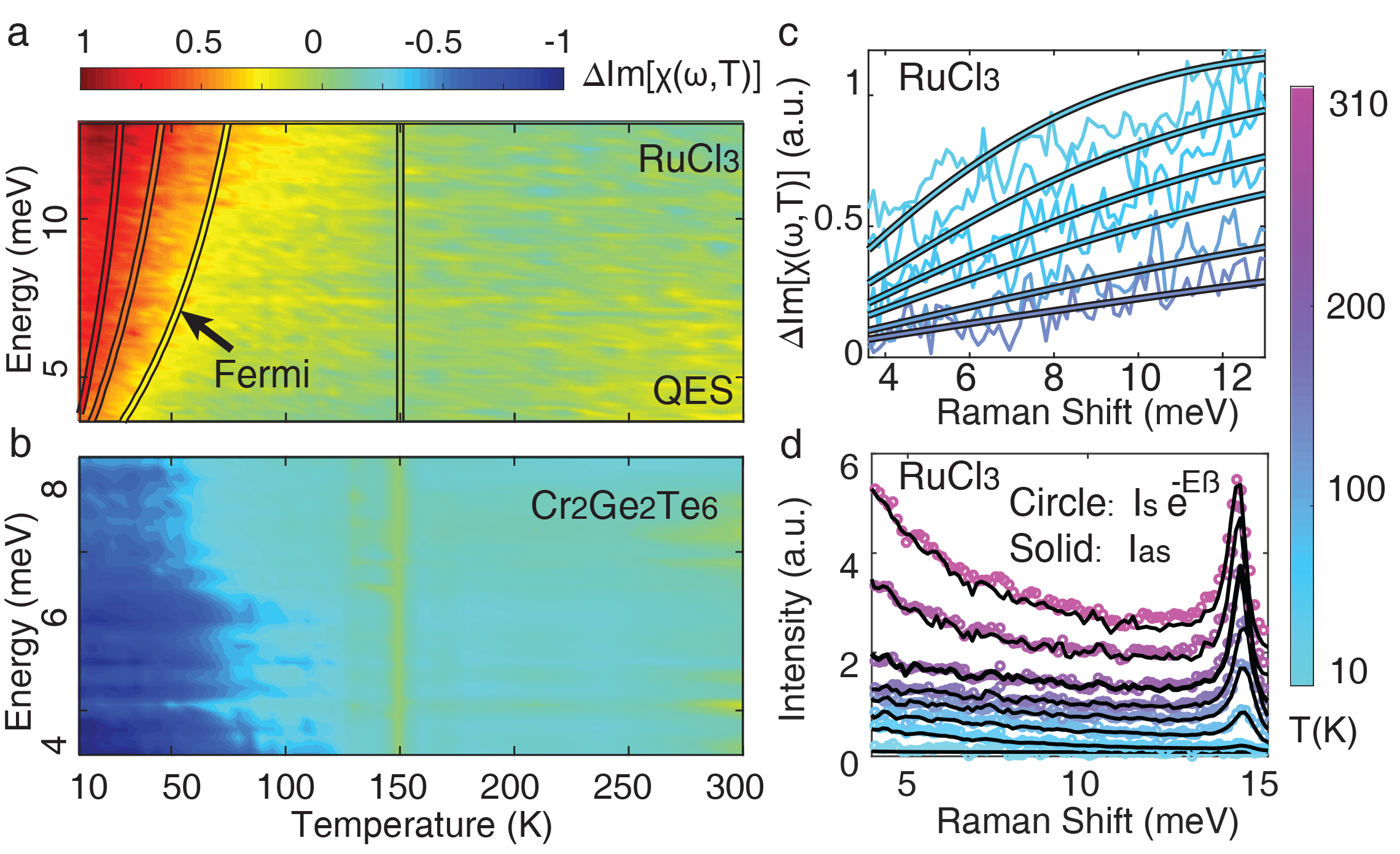}
    \caption{\textbf{The normalized Raman susceptibility and detailed balance} (\textbf{a}) Raman susceptibility of RuCl$_3$, $\Delta Im[\chi(\omega,T)] = Im[\chi(\omega,T)]-Im[\chi(\omega,150~K)]$). The curves with black outlines are the contour plot of Fermi functions ($\Delta n_{F}(\omega/2,T)=n_{F}(\omega/2,150)-n_{F}(\omega/2,T)$). Both data and the prediction are normalized to their maximum values. The agreement between the two confirms that Raman creates magnetic excitations that are made of pairs of fermions.  The upturn of the Raman intensity in the high temperature and low energy range results from thermal fluctuations of the magnetism (quasi-elastic scattering). (\textbf{b}) Raman susceptibility of a similar magnet, Cr$_{2}$Ge$_{2}$Te$_{6}$, where opposite to $\alpha$-RuCl$_{3}$, $\Delta Im[\chi(\omega,T)]$ is negative and does not match $n_{F}(\omega,T)$. (\textbf{c}) Comparison of $n_{F}(\omega,T)$ and $\Delta Im[\chi(\omega,T)]$ of RuCl$_3$ at fixed temperatures. The agreement further confirms the excitations are fermionic. (\textbf{d}) The excellent agreement between Stokes and anti-Stokes spectra of $\alpha$-RuCl$_{3}$ when normalized by the Boltzmann factor demonstrates the absence of laser heating.}
    \label{fig:Fig1}
\end{figure}

In inelastic light scattering, the measured intensity is determined by group theory, Fermi's golden rule, and via the fluctuation-dissipation theorem, related to the Raman susceptibility ($Im(\chi [\omega,T])$) times a Bose function\cite{devereaux:175}. In magnets this can produce peaks from single magnons, broad features reflecting the two-magnon joint density of states, or quasi-elastic scattering (QES) from thermal fluctuations\cite{Sandilands2015,Wulferding2010}. For the Kitaev QSL, Raman excites predominantly pairs of fractional particles in the energy range considered here($\approx 0.5J_K<\hbar\omega<\approx 2J_K$), leading to the energy loss ($I_S[\omega,T]$) and gain ($I_{aS}[\omega,T]$) intensities\cite{knolle2014raman,Nasu2016}:

\begin{align*}
I_S[\omega,T] = Im(\chi [\omega,T]) (n_B[\omega,T]+1) = JDos[\omega,T](1-n_F[\omega,T])^2\\
I_{aS}[\omega,T] = Im(\chi [\omega,T]) (n_B[\omega,T]) = JDos[\omega,T](n_F[\omega,T])^2
\end{align*}

where $n_{B/F}[\omega,T]$ are the Bose/Fermi distributions and $JDos[\omega,T]$ is approximately given by the joint density of states from the fractional particles.
\begin{sloppypar}
As shown in the supplemental, the fermionic response written above is consistent with the fluctuation dissipation theorem with the presence of time-reversal symmetry, requiring $I_{S}[\omega,T]/I_{aS}[\omega,T]=e^{\frac{\hbar\omega}{k_{B}T}}$\cite{devereaux:175,Misochko1996}.  This detailed balance analysis also allows us to confirm another crucial challenge in Raman measurements. Specifically, the extremely small Raman signals along with the low specific heat and thermal conductivity of $\alpha$-RuCl$_{3}$ suggest the required laser power may lead to significant heating. Indeed, we have found previous experiments likely suffered from significant laser induced heating, especially at low temperatures (see supplemental). One may suspect the apparent small upturn below 100 K in the Raman intensity of the continuum may only have been the result of laser induced heating instead of Fermi statistics. Furthermore, unless the temperature is well known, it is difficult to directly compare with the theoretical prediction for fractional statistics. In our current work we have made substantial improvement to the thermal anchoring and collection efficiency to allow for much lower laser fluency. In addition we have measured the anti-Stokes spectra such that we can employ detailed balance to check the actual temperature. In Fig.~\ref{fig:Fig1}d, we compare the anti-Stokes intensity and Stokes intensity times a Boltzmann factor with the measured temperature. The excellent agreement between them reveals that there is nearly no heating in the laser spot and thus we can use the measured crystal temperature. Unlike previous studies\cite{Sandilands2015,Glamazda2017}, our new measurement limits the possibility of laser heating to explain the low temperature upturn and confirms the sample is in detailed balance.
\end{sloppypar}
Returning to the statistics of the excitations in $\alpha$-RuCl$_{3}$, those governed by a Bose factor are best investigated via $Im(\chi[\omega,T])$\cite{devereaux:175,Misochko1996,Wulferding2010,Nakamura2015}. Using our new anti-Stokes spectra we are able to directly determine the Raman susceptibility from the difference between the Stokes and anti-Stokes intensities ($I_S-I_{aS} = Im(\chi[\omega,T])=JDos[\omega,T](1-2n_F(\omega,T))$). We explore the possibility that the susceptibility and thus the excitations are purely fermionic in Fig.~\ref{fig:Fig1}a, which removes the dark counts by showing the difference susceptibility: $\Delta Im(\chi[\omega,T])= Im(\chi[\omega,T]) - Im(\chi[\omega,150~K])$, and is limited to a range below the Raman active phonons\cite{Sandilands2016,Glamazda2017}. The utility of such an analysis is quite clear: namely the energy and temperature extent of the continuum can be directly observed - without contributions from high temperature quasi-elastic fluctuations or phonons. Beyond the continuum we can directly observe the crossover from spin liquid-like behavior (i.e. fractional continuum) to a standard paramagnet. Indeed, $\Delta Im(\chi[\omega,150~K\leq T\leq 200~K])$ is constant, as expected for a paramagnet. As discussed later, the response at high temperature is consistent with quasi-elastic scattering. Specifically, the Lorentzian at zero energy results from thermal fluctuations of the magnetism that confirm the magnetic specific heat is consistent with a standard paramagnet at high temperatures.

In addition to the continuum, these data directly confirm the presence of fractional excitations. We find excellent agreement between $\Delta Im(\chi[\omega,T\leq 150~K])$ and that expected for particles governed by Fermi statistics: $\Delta n_F[\omega/2,T]=n_F(\omega/2,150~K)-n_F(\omega/2,T)$, without the need to subtract any background from the quasi-elastic scattering. This is directly shown in  Fig.~\ref{fig:Fig1}c, which has constant temperature cuts of the data shown in Fig.~\ref{fig:Fig1}a, along with the calculated $\Delta n_{F}[\omega/2,T]$. The excellent agreement between the data and Fermi functions with half of the scattering energy confirms the presence of pairs of identical, fractional particles. This approach relies on a nearly energy and temperature independent $JDos[\omega,T]$, the presence of which we confirm later. To wit, this flat density of states of the fractional particles is consistent with numerical calculations for the Kitaev system at temperatures above the flux gap\cite{PhysRevB.92.115122}. We additionally performed the same analysis in another honeycomb system Cr$_2$Ge$_2$Te$_6$ (Fig.~\ref{fig:Fig1}b), grown by established methods and is ferromagnetic below $60~K$ with a similar Curie-Weiss temperature as $\alpha$-RuCl$_{3}$\cite{2013JAP...114k4907J}. The behavior of Cr$_2$Ge$_2$Te$_6$ is the exact opposite of $\alpha$-RuCl$_{3}$, namely, $\Delta Im(\chi[\omega,T])$ is negative throughout the whole measured range and decreases upon cooling.

\begin{figure}
    \centering
    \includegraphics[width=0.7\textwidth]{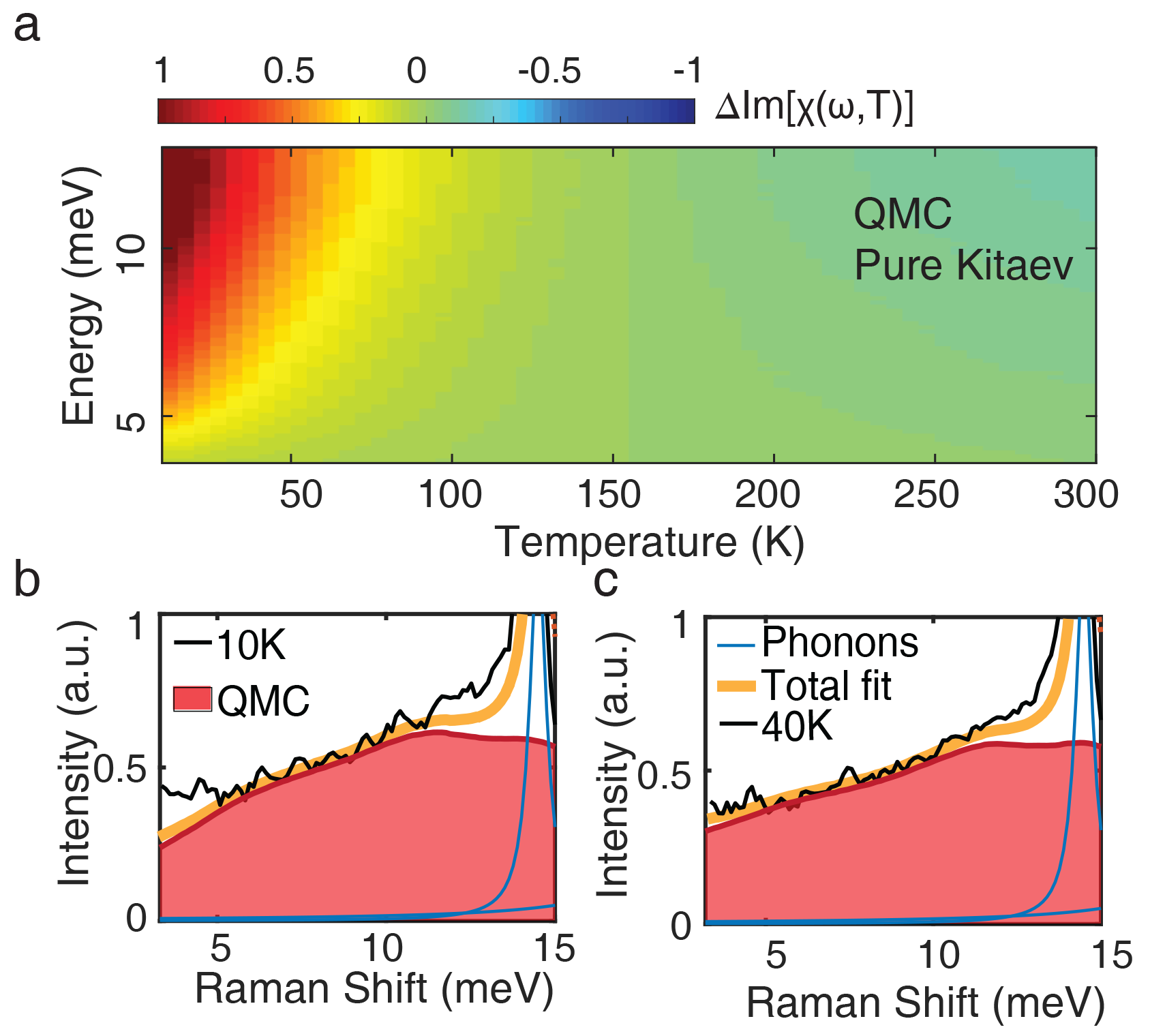}
    \caption{\textbf{Effects of non-Kitaev Terms} (\textbf{a}) Predicted Raman susceptibility for the pure Kitaev limit using QMC, where $\Delta Im[\chi(\omega,T)] = Im[\chi(\omega,T)]-Im[\chi(\omega,150~K)]$).  (\textbf{b}) The measured intensity in XX polarization of $\alpha$-RuCl$_{3}$ (black line) compared with the calculated result of the pure Kitaev limit (shaded red) and a Lorentzian phonon (solid blue). The total fit (bold orange) deviates at low energies for the 10 K data, due to non-Kitaev terms. (\textbf{c}) By 40 K there is nearly perfect agreement, which indicates the non-Kitaev terms are not relevant in this energy and temperature range.}
    
    \label{fig:Fig2}
\end{figure}

Beyond establishing the presence of fractional particles, our new data set provides additional insights into the importance of non-Kitaev terms in $\alpha$-RuCl$_{3}$. Indeed, it is somewhat surprising to find fractional response in a material known to order anti-ferromagnetically at low temperatures. This order has been attributed to the presence of additional symmetry allowed terms, such as Heisenberg or off-diagonal exchange.\cite{PhysRevB.93.155143,Winter2017,Catuneanu2018} To better understand the role of Kitaev versus other terms, we calculated the Raman response in the pure Kitaev limit using quantum Monte Carlo (QMC). The resulting Raman susceptibility for a pure Kitaev model is shown in Fig.~\ref{fig:Fig2}a, which is quite similar to the measured data (Fig.~\ref{fig:Fig1}a) at high energy and/or temperature. To confirm this we directly compare the QMC and the Raman response at 10 and 40 K in  Fig.~\ref{fig:Fig2}b and c. To account for the contribution of the 15 meV phonon we add a Lorentzian fit of the phonon to the QMC calculation. The total fit shows very good overlap with the spectra between 6 to 13 meV at 10 K, and excellent agreement at all measured energies for 40 K. At low temperatures and energies the measured Raman intensity is larger than the prediction of the QMC. We attribute this discrepancy to the presence of the additional, weaker terms, consistent with recent exact diagonalization calculations.\cite{Rousochatzakis2018} As such, our combined theoretical and new Raman results confirm the presence of fractional excitations on short time scales (i.e. high energies) in $\alpha$-RuCl$_{3}$, likely from its proximity to a quantum spin liquid state. 
 
\begin{figure}
    \centering
    \includegraphics[width=1\textwidth]{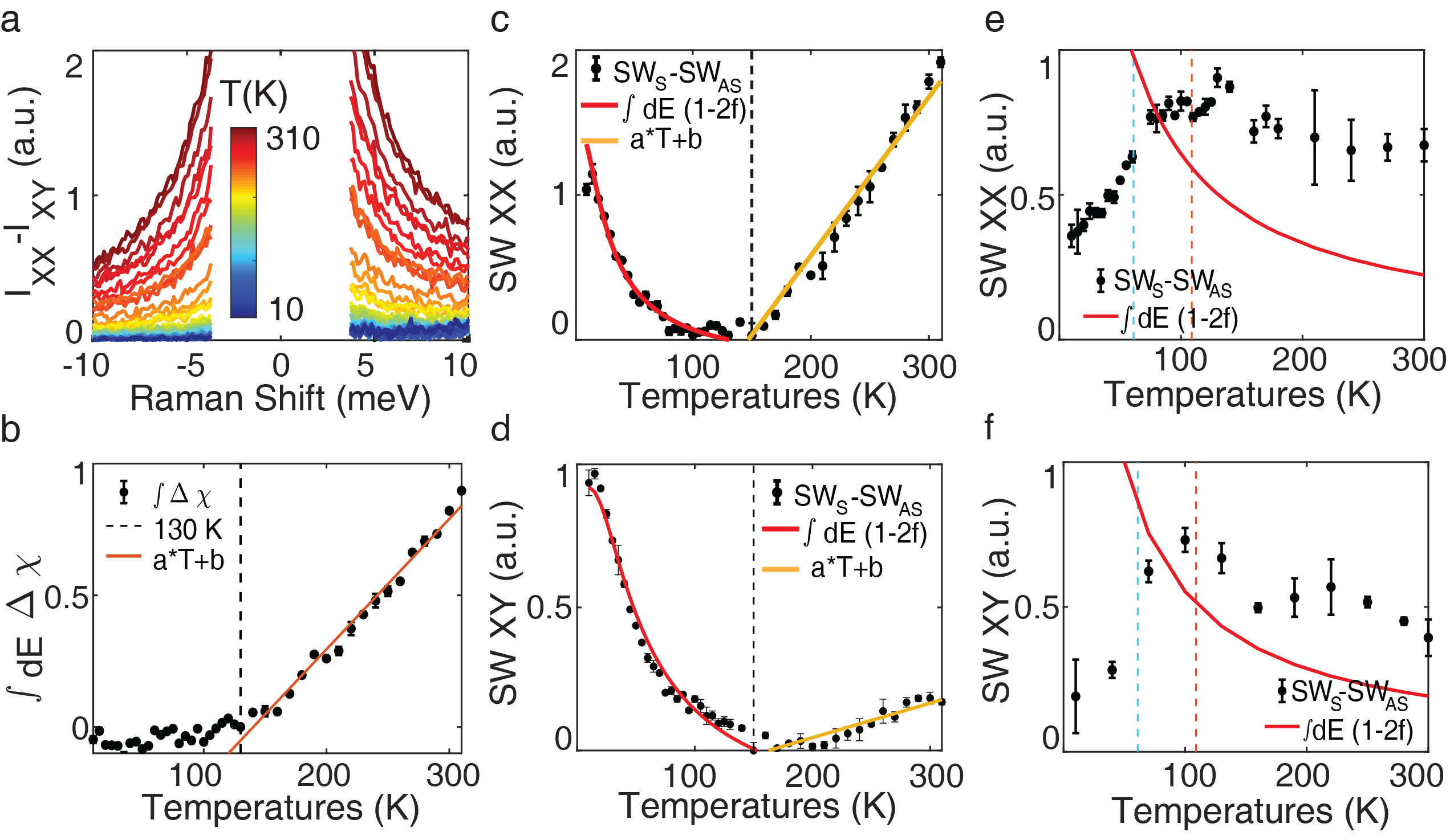}
    \caption{\textbf{Limit of Fermi statistics} (\textbf{a}) The continuum in $\alpha$-RuCl$_{3}$ due to fractional particles is removed by taking the difference between XY and XX intensities. This confirms the continuum is consistent with predictions of the Kitaev model, and the high temperature response is from quasi-elastic scattering (i.e. Lorentzian times a Bose factor). (\textbf{b}) The integration of the Raman susceptibility with only the quasi-elastic scattering response, reveals a linear T behavior above 150 K and temperature independent behavior below. (\textbf{c} \& \textbf{d}) Integrated spectral weight of $Im[\chi(\omega,T)]$, reveals Fermi statistics in $\alpha$-RuCl$_{3}$ below $\approx 100$ K (solid red line) in XX and XY  polarizations. Above $130$ K the response is linear in temperature due to the quasi-elastic scattering (yellow lines). The spectral weight from Cr$_{2}$Ge$_{2}$Te$_{6}$ (\textbf{e} \& \textbf{f}) is enhanced up to $T_{C}$ (blue dashed line) but the temperature dependence above does not fit that expected for fermions (solid red line).}
    \label{fig:Fig3}
\end{figure}

To ensure our approach is self-consistent it is highly desirable to also analyze the integrated Raman response, as done previously in $\alpha$-RuCl$_{3}$ and Li$_{2}$IrO$_{3}$.\cite{Nasu2016,Sandilands2015,Glamazda2017} Nonetheless, it is also crucial to find a reliable method to separate the quasi-elastic response from the continuum such that it can be independently studied and further confirm the presence of Fermi statistics. This is now possible using both the polarization and Stokes minus anti-Stokes spectra ($Im[\chi[\omega,T]]$). Since the continuum has equal weight in both polarizations\cite{knolle2014raman,Sandilands2015} it can be removed via their difference: $\Delta I_{S/aS}[\omega,T]=I^{XX}_{S/aS}[\omega,T]-I^{XY}_{S/aS}[\omega,T]$. As seen in Fig.~\ref{fig:Fig3}a, $\Delta I_{S/aS}[\omega,T]$ is consistent with thermal fluctuations (i.e. QES)\cite{Wulferding2010,Nakamura2015}, namely a Lorentzian whose amplitude is given by the magnetic specific heat ($C_{m}[T]$) times temperature and appropriately weighted Bose factors (i.e. greater Stokes than anti-Stokes intensity). Next we determined the QES amplitude via the spectral weight (SW) of the Raman susceptibility: $SW_{QES}[T]=\int \chi^{QES}_{XX}[\omega,T]-\chi^{QES}_{XY}[\omega,T]d\omega=\int dE\Delta \chi$. Consistent with direct fits of the $\Delta I_{S/aS}[\omega,T]$ (see supplemental) and robust to the limits of integration (as long as phonons are not included), we find $SW_{QES}[T]\propto T$ (see Fig.~\ref{fig:Fig3}b). This suggests the magnetic specific heat is temperature independent, as expected for a classical paramagnet at high temperatures. Since the QES signal is nearly zero in $\chi[\omega,T<150~K]$, this confirms the Raman susceptibility (and not the intensity) naturally separates the QES from the continuum. Thus our new measurements reveal the energy/temperature range over which the excitations are fractional without contributions from other bosonic modes.
 
Having isolated the QES and found its temperature dependence, we determine the temperature bounds of the Fermi statistics. Specifically, we investigate the difference between the Stokes and anti-Stokes SW in a given polarization ($\Delta SW[T]=\int (I_{S}[\omega,T]-I_{aS}[\omega,T])d\omega$), which includes the integrated Fermi function from the fractional excitations and the QES contribution (see supplemental). As shown in Fig.~\ref{fig:Fig3}c \& d for two different polarizations, the integrated weight follows the expected response for pairs of fermionic excitations until $T\approx 150~K$ where it crosses over to a linear temperature dependence from the QES. The fermionic response is equal in both polarizations, consistent with the Kitaev model\cite{knolle2014raman}. Thus with just three parameters, one fixed by the lowest temperature, we fully explain the SW for all energy ranges, temperatures, and polarizations. To further confirm this, we tried the same analysis on our new Cr$_2$Ge$_2$Te$_6$ data. As shown in Fig.~\ref{fig:Fig3}e\& f the difference between the Stokes and anti-Stokes of Cr$_2$Ge$_2$Te$_6$ cannot be fit with a Fermi function at all. Thus the results presented in Fig.~\ref{fig:Fig3}c,d provide a quantitative confirmation of the presence of fractional excitations up to high temperatures.

\begin{figure}
    \centering
    \includegraphics[width=1\textwidth]{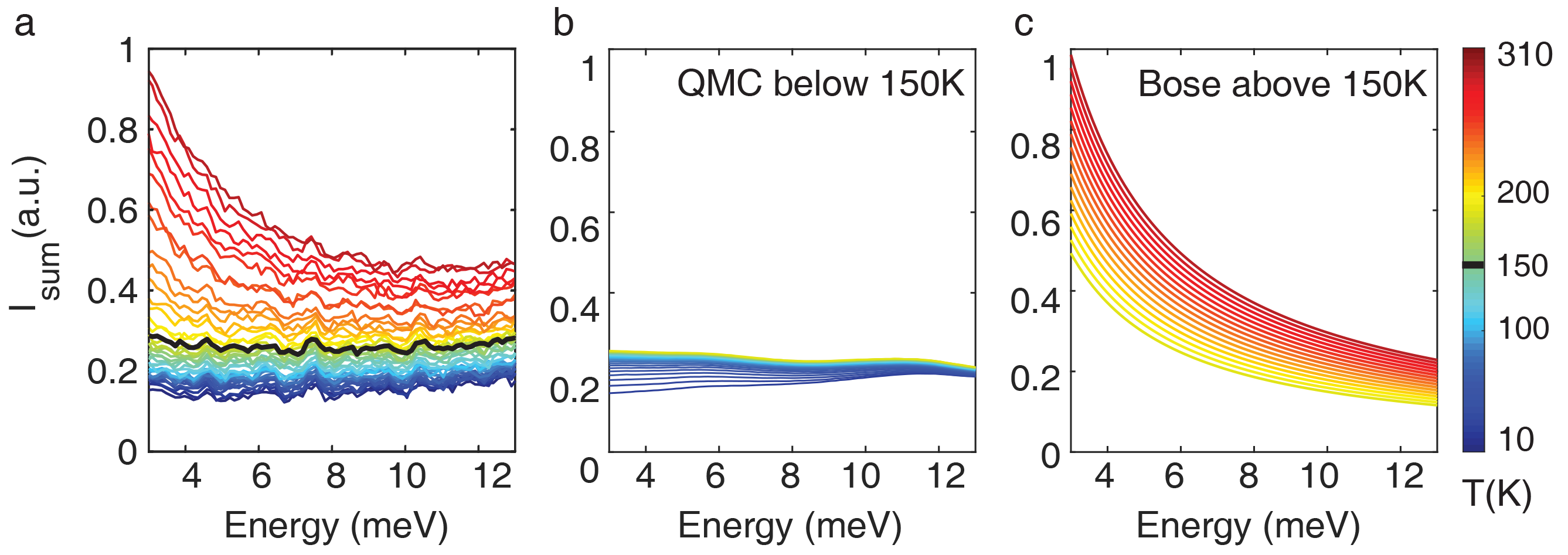}
    \caption{\textbf{Joint Density of States} (\textbf{a}) The joint density of states, assuming the response is governed by fermi statistics. Plotted is I$_{sum}$[$\omega$, T ], the sum of the loss and gain intensity, which is nearly temperature independent for $T < 150 K$. This is consistent with the QMC (\textbf{b}) results for the Kitaev model. For the full temperature range, $I_{sum}$ reveals an upturn from thermal fluctuations, which is consistent with the pure Bose function calculation (\textbf{c}).}
    \label{fig:Fig4}
\end{figure}

To further confirm these Fermi signatures we return to the expected Stokes and anti-Stokes Raman intensities. So far we focused on the difference between Stokes and anti-Stokes as it removes the Bose factor. However for fermions obeying the Pauli exclusion principle we expect their sum to be conserved. Since Raman creates pairs of fermions, we find that if we add the square root of the Stokes intensity with the square root of the anti-Stokes intensity, the Fermi factor is removed. Thus we introduce the quantity $I_{sum}[\omega,T] = (\sqrt{I_S[\omega,T]}+\sqrt{I_{aS}[\omega,T]})^2=JDos[\omega,T]$ (see Fig.~\ref{fig:Fig4}a ). The resulting quantity provides both an important check of the Fermi statistics and new information on the joint density of states of the fractional particles that has so far been elusive. We find the low temperature $I_{sum}[\omega,T<150~K]$ is relatively temperature and energy independent over almost an order of magnitude of Raman shift ($3.6~meV < \hbar\omega < 12~meV$), consistent with QMC prediction in Fig.~\ref{fig:Fig4}b. Thus our results suggest that at temperatures well above the flux gap, $\alpha$-RuCl$_{3}$ behaves as a Majorana metal with a featureless density of states. 

The sum of the responses at low temperatures is in stark contrast to the high temperature bosonic response resulting from the crossover to a normal paramagnet. As shown in Fig.~\ref{fig:Fig4}a, $I_{sum}[\omega,T>150K]$ displays a low energy divergence that grows with heating due to the thermal fluctuation of the magnetic system (QES). This is consistent with the predicted response for bosons (Fig.~\ref{fig:Fig4}c), where a constant Raman susceptibility is assumed. As expected for particles without the Pauli-exclusion principle, there is no conservation of particle number and a strong enhancement of the signal.

The Raman susceptibility, rather than the intensity of the Stokes and anti-Stokes spectra (Figs.~\ref{fig:Fig1}), provides direct evidence that the magnetic excitations in $\alpha$-RuCl$_{3}$ are fractional, following Fermi statistics. Alternatively, if Fermi statistics is assumed, the extracted joint density of states, shown in Fig. \ref{fig:Fig4}, is nearly flat and temperature independent. At higher temperatures and energies, these results are consistent with calculations in the exact Kitaev limit, deviating only at low energy and temperature due to additional terms. Thus our results provide concrete evidence for the fractional nature of the excitations at short time scales, due to proximity to a QSL state. We note this is also at energies and temperatures above the flux gap, suggesting the fractional excitations and their Fermi statistics are robust on shorter time scales than those associated with other terms of fluxes. By comparing with QMC calculations we suggest the energy and temperature bounds of non-Kitaev terms in $\alpha$-RuCl$_{3}$. We thus establish the Raman susceptibility as a new standard for determining the fractional nature of the excitations. As such the measurement of the Raman susceptibility can be used to probe the transition with external perturbations into the pure spin liquid state.

\newpage
\begin{methods}

\noindent
\textbf{RuCl$_{3}$ crystal growth, handling and characterization.}
Single crystals of $\alpha$-RuCl$_{3}$ were prepared using high-temperature vapor-transport techniques from pure $\alpha$-RuCl$_{3}$ powder with no additional transport agent. Crystals grown by an identical method have been extensively characterized via bulk and neutron scattering techniques\cite{Cao2016,Banerjee2017,Banerjee2018} revealing behavior consistent with what is expected for a relativistic Mott insulator with a large Kiteav interaction\cite{PLumb2014,PhysRevB.91.155135,Sandilands2016,Koitzsch2016,Zhou2016,Yadav2016,Do2017,Hirobe2017,Winter2017,Little2017,Jansa2018,PhysRevB.93.155143,Leahy:2017cv,Wellm2018}. The crystals have been shown to consistently exhibit a single dominant magnetic phase at low temperature with a transition temperature $T_{N}\approx 7$ K, indicating high crystal quality with minimal stacking faults\cite{Cao2016}. Care was taken in mounting the crystals to minimize the introduction of additional stacking faults, as evidenced by the high reproducibility of the spectra across different crystals and experimental setups. Characterization was consistent with previous studies\cite{PLumb2014,PhysRevB.91.180401,WangLoidl2017,Zheng2017}.

\noindent
\textbf{Raman spectroscopy experiments.}
Since Raman scattering involves a photon in and photon out, it allows one to measure both the symmetry and energy change of an excitation. Furthermore, one can choose an energy and/or symmetry channel to separate the magnetic, electronic and lattice responses\cite{Nasu2016,Sandilands2015,devereaux:175,Wulferding2010,Nakamura2015,YaoCGT2Dmat,devereaux:175,Glamazda2017}. The majority of the Raman experiments were performed with a custom built, low temperature microscopy setup\cite{Tian:2016ja}. A 532 nm excitation laser, whose spot has a diameter of 2 $\mu m$, was used with the power limited to 30 $\mu$W to minimize sample heating while allowing for a strong enough signal. The sample was mounted by thermal epoxy onto a copper \emph{xyz} stage. At both room and base temperature the reported spectra were averaged from three spectra in the same environment to ensure reproducibility. The spectrometer had a 2400 g/mm grating, with an Andor CCD, providing a resolution of $\approx 1$ cm$^{-1}$. To minimize the effects of hysteresis from the crystal structural transition, data was taken by first cooling the crystal to base temperature, and once cooled to base temperature, spectra were acquired either every 5 or 10 K by directly heating to that temperature. The absence of hysteresis effects was confirmed by taking numerous spectra at the same temperature after different thermal cycles (100 K in the middle of the hysteresis region). In addition, recent studies of the Raman spectra of RuCl$_{3}$ suggest an effect of the surface structure upon exposure to air\cite{Zhou2018,mashhadi2018electrical}. To minimize this, crystals were freshly cleaved and immediately placed in vacuum within three minutes. Moreover, a recently developed wavelet based approach was employed to remove cosmic rays\cite{TianSpike}.

\noindent
\textbf{Quantum Monte Carlo Calculations.} 
The Hamiltonian of the Kitaev model on the honeycomb lattice is given by
 \begin{equation}
  {\cal H}=-J_x\sum_{\means{jk}_x}S_j^x S_k^x-J_y\sum_{\means{jk}_y}S_j^y S_k^y-J_z\sum_{\means{jk}_z}S_j^z S_k^z,
  \label{eq:H_spin}
 \end{equation}
 where $\bm{S}_j$ represents an $S=1/2$ spin on site $j$, and $\means{jk}_\gamma$ stands for a nearest-neighbor (NN) $\gamma(=x,y,z)$ bond shown in Fig.~1a.
In the calculation for the spectrum of the Raman scattering we adopt the Loudon-Fleury (LF) approach.
The LF operator for the Kitaev model is given by
\begin{equation}
 {\cal R}=\sum_{\means{ij}_\alpha}(\bm{\epsilon}_{\rm in}\cdot \bm{d}^\alpha)(\bm{\epsilon}_{\rm out}\cdot \bm{d}^\alpha)J_\alpha S_i^\alpha S_j^\alpha,\label{eq:LF}
\end{equation}
where $\epsilon_{\rm in}$ and $\epsilon_{\rm out}$ are the polarization vectors of the incoming and outgoing photons and $\bm{d}^\alpha$ is the vector connecting a NN $\alpha$ bond\cite{knolle2014raman,fleury1968scattering}.
Using this LF operator, the Raman spectrum is calculated as
\begin{equation}
 I(\omega)=\frac{1}{N}\int_{-\infty}^{\infty}dt e^{i\omega t}\means{{\cal R}(t){\cal R}},\label{eq:raman}
\end{equation}
where ${\cal R}(t)=e^{i{\cal H}t}{\cal R}e^{-i{\cal H}t}$ is the Heisenberg representation.
The temperature dependence of $I(\omega)$ is numerically evaluated using the Monte Carlo simulation in the Majorana fermion representation without any approximation\cite{Nasu2014}.
In the following we show the details of the calculation procedure\cite{Nasu2016}.

Using the Jordan-Wigner transformation, the Hamiltonian is mapped onto the Majorana fermion model as
\begin{equation}
   {\cal H}=\frac{iJ_x}{4}\sum_{(jj')_x}c_j c_k-\frac{iJ_y}{4}\sum_{(jj')_y}c_j c_k-\frac{iJ_z}{4}\sum_{(jj')_z} \eta_r c_j c_k,
\end{equation}
 where $(jj')_{\gamma}$ is the NN pair satisfying $j<j'$ on the $\gamma$ bond, and $\eta_{r}$ is a $Z_{2}$ conserved quantity defined on the $z$ bond ($r$ is the label for the bond), which takes $\pm 1$.
This Hamiltonian is simply written as
\begin{equation}
 {\cal H}=\frac{1}{2}\sum_{jk}A_{jk}(\{\eta_r\}) c_j c_k,
\end{equation}
using the Hermitian matrix $A_{jk}(\{\eta_{r}\})$ depending on the configuration of $\{\eta_{r}\}$.
The LF operator shown in Eq.~(\ref{eq:LF}) is also given by the bilinear form of the Majorana fermion:
\begin{equation}
 {\cal R}(\{\eta_r\})=\frac{1}{2}\sum_{jk}B_{jk}(\{\eta_r\}) c_j c_k,
\end{equation}
where $B(\{\eta_r\})$ is a Hermitian matrix.
To evaluate Eq.~(\ref{eq:raman}), we separate the sum over the states into $\{c_j\}$ and $\{\eta_r\}$ parts:
\begin{equation}
 I(\omega)= \frac{1}{Z}\sum_{\{\eta_r=\pm1\}}\bar{I}(\omega;\{\eta_r\})e^{-\beta F_f(\{\eta_r\})},
  \end{equation}
with
\begin{equation}
 \bar{I}(\omega;\{\eta_r\})=\frac{1}{Z_f(\{\eta_r\})}{\rm Tr}_{\{c_j\}}\left[\frac{1}{N}\int_{-\infty}^{\infty}dt e^{i\omega t}{\cal R}(t;\{\eta_r\}){\cal R}(\{\eta_r\})e^{-\beta {\cal H}(\{\eta_r\})}\right],\label{eq:raman_eta}
\end{equation}
where $Z=\sum_{\{\eta_r=\pm1\}}e^{-\beta F_f(\{\eta_r\})}$ and $Z_f(\{\eta_r\})=e^{-\beta F_f(\{\eta_r\})}={\rm Tr}_{\{c_j\}}e^{-\beta {\cal H}(\{\eta_r\})}$.
By applying Wick's theorem to Eq.~(\ref{eq:raman_eta}), we calculate the Raman spectrum at $\omega (\neq 0)$ for a given configuration $\{\eta_r\}$ as
\begin{align}
 \bar{I}(\omega;\{\eta_r\})=&\frac{1}{N}\sum_{\lambda\lambda'}\Bigl[
 2\pi|C_{\lambda\lambda'}|^2 f(\varepsilon_\lambda)[1-f(\varepsilon_{\lambda'})]\delta(\omega+\varepsilon_\lambda-\varepsilon_{\lambda'})\notag\\
 &+\pi|D_{\lambda\lambda'}|^2[1-f(\varepsilon_\lambda)][1-f(\varepsilon_{\lambda'})]\delta(\omega-\varepsilon_\lambda-\varepsilon_{\lambda'})\notag\\
 &+\pi|D_{\lambda\lambda'}|^2f(\varepsilon_\lambda)f(\varepsilon_{\lambda'})\delta(\omega+\varepsilon_\lambda+\varepsilon_{\lambda'}) \Bigr],
  \end{align}
where $f(\varepsilon)=1/(1+e^{\beta\varepsilon})$ is the Fermi distribution function with zero chemical potential, $\{\varepsilon_\lambda\}$ is the set of the positive eigenvalues of $A$ with the eigenvectors $\{\bm{u}_{\lambda}\}$, and the matrices $C$ and $D$ are given by $C_{\lambda\lambda'}=2 \bm{u}_{\lambda}^\dagger B \bm{u}_{\lambda'}$ and $D_{\lambda\lambda'}=2 \bm{u}_{\lambda}^\dagger B \bm{u}_{\lambda'}^*$.
In the Monte Carlo simulations, we generate a sequence of configurations of $\{\eta_r\}$ to reproduce the distribution of $e^{-\beta F_f(\{\eta_r\})}$, and hence the finite-temperature spectrum is simply computed as $I(\omega)=\means{\bar{I}(\omega;\{\eta_r\})}_{\rm MC}$ with $\means{\cdots}_{\rm MC}$ being the Monte Carlo average.

\noindent
\textbf{Correction for optical constants.}
According to the Beer-Lambert Law, the intensity of the laser decreases exponentially with the depth:
$I[z]=I_0 e^{-\alpha z}$, where $d$ is the depth and $\alpha$ is the attenuation constant, which is a function of laser frequency and dielectric constant of the material ($\alpha =\frac{\omega}{c}Im[\tilde{n}(\omega)] =-\frac{4\pi E[\omega_{0}]}{hc}k[\omega_{0}]$). Alternatively one can express this in terms of a penetration depth indicating the length scale relevant to absorption: $\delta=\frac{1}{\alpha}$. Applying this to our experiment, for a certain depth $d$, we find the incident laser intensity as a function of distance from the surface, $I_{in}[\omega_{0},z]=I_{0} e^{-\frac{4\pi E[\omega_{0}]}{hc}k[\omega_{0}]z}$. Here, $\omega_{0}$ is the frequency of the excitation laser, $I_{0}$ is the initial incoming laser power in front of the sample,and $\delta$($\approx$140 nm) is much shorter than the thickness of $\alpha$-RuCl$_{3}$ bulk crystal. 
To properly account for the temperature dependence of the optical constants on the measured Raman signal, it is crucial to account for these absorption losses. Specifically the measured intensity is reduced by the absorption of the outgoing Raman photons, (i.e. $I_{out}[\omega,\omega_{0},z]=I_{in}[\omega_{0},z]e^{-\frac{4\pi E[\omega]}{hc}k[\omega]z}$) 
where $\omega$ is the frequency of the scattered light. Furthermore, one should also consider the probability of transmission at the surface of $\alpha$-RuCl$_{3}$ ($T[\omega]$), which also depends on the Raman light frequency. Applying the transmission rate to the Raman signal, we obtain the Raman intensity coming out of the sample at each point $I_{Raman}[\omega,\omega_{0},z] =I_{out}[\omega,\omega_{0},z] *T[\omega]$. Finally, one obtains the signal intensity by integrating the attenuated intensity of scattering point at each depth via $I_{corrected}[\omega_{0},\omega]=\int_{0}^{d_{max}} I_{Raman}[\omega,\omega_{0},z] dz $\cite{PhysRevB.98.014308,Tian:2016ja,Yoon:2009p2241}.
All presented Raman data in this paper are corrected by this method using the previously published optical constants\cite{sandilands2016optical}. 
\end{methods}

\begin{addendum}
\item [Acknowledgements] We are grateful for numerous discussions with Natalia Perkins, Joshua Heath, Kevin Bedell and Ying Ran. The Raman experiments at 532~nm were performed by Y.W. with support from the National Science Foundation, Award No. DMR-1709987. G.B.O. assisted in the analysis with support from the U.S. Department of Energy (DOE), Office of Science, Office of Basic Energy Sciences under Award No. DE-SC0018675. Raman experiments performed at 720~nm (T.G. and J.Y.) were achieved by support from the National Science Foundation, Award No. ECCS 1509599. The crystal growth and characterization of $\alpha$-RuCl$_{3}$ (P.L.K. and D.M.) with , while A.B. and S.N. were supported by the US DOE Basic Energy Sciences Division of Scientific User Facilities. The work on Cr$_{2}$Ge$_{2}$Te$_{6}$ (H.J. and R.J.C.) at Princeton University is sponsored by an ARO MURI, grant W911NF1210461. The numerical simulations were performed by J.N. with support from Grants-in-Aid for Scientific Research (KAKENHI) (numbers JP15K13533, JP16H02206, JP16K17747, and JP18H04223). Parts of the numerical calculations were performed in the supercomputing systems in ISSP, the University of Tokyo.

\item[Competing Interests] The authors declare that they have no competing financial interests.

\item[Author Contributions] Y.W. performed the Raman experiments, with assistance from G.B.O., T.G., and J.Y. Analysis was done by Y.W., J.N., J.K., and G.B.O. The crystal growth and initial characterization were done by P.L., A.B., and D.M. The design of the experiments and conception of the study were achieved by K.S.B, S.N. and J.K. The theoretical calculations were performed by J.N., J.K. and Y.M.  

\item[Correspondence] Correspondence and requests for materials
should be addressed to K.S. Burch~(email: ks.burch@bc.edu).

\end{addendum}

\bibliography{References}

\end{document}